# *ASB1* differential methylation in ischaemic cardiomyopathy. Relationship with left ventricular performance in end stage heart failure patients

Ana Ortega (1)[¶], Estefanía Tarazón (1)[¶], Carolina Gil-Cayuela (1), Luis Martínez-Dolz (2), Francisca Lago (3), José Ramón González-Juanatey (3), Juan Sandoval (4), Manuel Portolés (1), Esther Roselló-Lletí (1)[&], Miguel Rivera (1)[&].

(1) Cardiocirculatory Unit, Health Research Institute La Fe, Valencia, Spain. (2) Heart Failure and Transplantation Unit, Cardiology Department, University and Polytechnic La Fe Hospital, Valencia, Spain. (3) Cellular and Molecular Cardiology Research Unit, Department of Cardiology and Institute of Biomedical Research, University Clinical Hospital, Santiago de Compostela, Spain. (4) Epigenomic Unit, Health Research Institute La Fe, Valencia, Spain. [¶]These authors contributed equally to this work. [&]These authors also contributed equally to this work.

**Aims:** Ischaemic cardiomyopathy (ICM) leads to impaired contraction and ventricular dysfunction causing high rates of morbidity and mortality. Epigenomics allows the identification of epigenetic signatures in human diseases. We analyse the differential epigenetic patterns of *ASB* gene family in ICM patients and relate these alterations to their haemodynamic and functional status.

**Methods and Results:** Epigenomic analysis was carried out using 16 left ventricular (LV) tissue samples, 8 from ICM patients undergoing heart transplantation and 8 from control (CNT) subjects without cardiac disease. We increased the sample size up to 13 ICM and 10 CNT for RNA-sequencing and to 14 ICM for pyrosequencing analyses. We found a hypermethylated profile (cg11189868) in the *ASB1* gene that showed a differential methylation of 0.26Δβ, $P < 0.05$. This result was validated by pyrosequencing technique (0.23Δβ, $P < 0.05$). Notably, the methylation pattern was strongly related to LV ejection fraction ($r = -0.849$, $P = 0.008$) stroke volume ($r = -0.929$, $P = 0.001$) and end-systolic and diastolic LV diameters ($r = -0.743$, $P = 0.035$ for both). *ASB1* showed a down regulation in mRNA levels (-1.2 fold, $P < 0.05$).

**Conclusion:** Our findings link a specific *ASB1* methylation pattern to LV structure and performance in end-stage ICM, opening new therapeutic opportunities and providing new insights regarding which is the functionally relevant genome in the ischemic failing myocardium.

# INTRODUCTION

There are several large-scale studies deciphering the alterations in human heart failure (HF) proteome and genome, to elucidate the molecular mechanisms involved in the pathophysiology of this syndrome.[1, 2] However, apart from these global changes, there are other omic approaches revealing insights in the regulation of expression patterns in disease. Epigenomics has allowed the identification of epigenetic signatures in human diseases including different types of cancer,[3-5] neurological disorders[6, 7] or infections.[8] There are also some evidences of their role in cardiovascular diseases.[9-11] Moreover, the development of the novel approach MethylationEPIC BeadChip (Infinium) microarray, has improved greatly the coverage, reaching 850,000 CpG methylation sites (850K).[12]

Our group has extensively studied different pathophysiological factors in HF, including apoptosis processes[13] and cytoskeletal and cell adhesion molecular alterations[14, 15] that are important to clinical status of patients. Interestingly, the gene family *ASB* codifies Ankyrin repeat and SOCS box proteins that mediate, through their specific domains, protein-protein interactions, protein synthesis, and myogenesis and proteasomal degradation processes.[16-18] Moreover members of this family have been related to skeletal muscle mass regulation.[19]

Taking into account these previous data and the increasing evidence of the influence of epigenetic changes in the pathophysiology of human diseases, we hypothesize that the epigenetic changes in the *ASB* gene family may relate to the ischaemic left ventricular (LV) performance. We analysed specific gene methylation changes in patients with end stage ischaemic cardiomyopathy (ICM) compared with those in control (CNT) subjects. Furthermore, we related these alterations with LV function and with invasively calculated stroke volume (SV).

**METHODS**

**Cardiac tissue samples**

Epigenomic experiments were conducted with 8 LV tissue samples from ICM patients undergoing cardiac transplantation and 8 non-diseased donor hearts (CNT), increasing the sample size for RNA-sequencing (13 ICM) and pyrosequencing (14 ICM). Clinical history, hemodynamic study, ECG, and Doppler echocardiography data were available. These data were collected by physicians blind to the subsequent analysis of LV function that was carried out. Patients were functionally classified according to the NYHA criteria and were receiving medical treatment following the guidelines of the European Society of Cardiology.[20]

All CNTs had normal LV function (LVEF>50%) and none had any history of cardiac disease. Samples were obtained from non-diseased donor hearts that had been rejected for cardiac transplantation owing to size or blood type incompatibility. Donors died of either cerebrovascular or motor vehicle accidents.

Tissue samples were collected from near the apex of left ventricle, maintained in 0.9% NaCl, and preserved at 4 °C for a maximum of 4.4±3 h after the coronary circulation loss. Then, were stored at −80°C until used. Appropriate handling and rapid sample collection and storage, by our on call (24 h) team over the last 10 years, led to the obtaining of high quality samples (RIN≥9 and DNA ratios 260/280 ~1.8 and 260/230 ~2.0).

This study was approved by the Ethics Committee (Biomedical Investigation Ethics Committee of La Fe University Hospital of Valencia, Spain), and was conducted in accordance with the guidelines of the Declaration of Helsinki.[21] Signed informed consent was obtained from each patient.

**DNA extraction, quality assessment and Infinium MethylationEPIC BeadChip**

DNA was extracted using a modified phenol-chloroform protocol. The epigenomic study (Infinium MethylationEPIC BeadChip platform (Illumina) used the HumanMethylation450 BeadChip protocol. Methylation level was displayed as beta values ranging from 0–1. Beta-values with detection $P>0.01$ were removed from the analysis. The raw data were normalized and background corrected. The resulting raw data (IDATs) were normalized and background corrected using the methylation module (1.9.0) in GenomeStudio (v2011.1) software.

**Validation by pyrosequencing**

500ng of DNA were converted using the EZ DNA Methylation Gold (Zymo Research) bisulfite conversion kit, following the manufacturer's recommendations. Specific sets of primers for PCR amplification and sequencing were designed (PyroMark assay design v2.0.01.15) to hybridize with CpG-free sites ensuring methylation-independent amplification. PCR was performed with biotinylated primers and the PyroMark Vacuum Prep Tool (Biotage, Sweden) was used to prepare single-stranded PCR products. Reactions were performed in a PyroMark Q24 System version 2.0.6 (Qiagen) and the methylation value was obtained from the average of the CpG dinucleotides included in the sequence analysed.

**RNA-sequencing and computational analysis**

These protocols were performed as described in.[14] The data presented in this manuscript have been deposited in NCBI's Gene Expression Omnibus (GEO)[22] (GSE55296) (http://www.ncbi.nlm.nih.gov/geo/query/acc.cgi?acc=GSE55296).

**Statistical methods**

Data were analysed as previously described.[14] $P<0.05$ was considered statistically significant. CpGs with $\Delta\beta \geq \pm 0.1$ were considered differentially methylated.

## RESULTS

### Clinical characteristics of patients

For epigenomic studies we analyzed 8 ICM samples, patients were all men with a mean age of 53±5 years, and 8 CNT samples, 63% men with a mean age of 59±20 years. Sample size was increased for RNA-sequencing (13 ICM), all men with a mean age of 54±7 years and 10 CNT, 80% men with a mean age of 47±16 years. And also for pyrosequencing (14 ICM), all men with a mean age of 53±6 years. Patients had a NYHA classification of III-IV and were diagnosed with comorbidities including hypertension and diabetes mellitus. Comorbidities and other echocardiographic data were not available for the CNT group, in accordance with the Spanish Organic Law on Data Protection 15/1999. Clinical characteristics of patients are shown in Table 1.

### Methylation profile of *ASB* family and gene expression analysis of *ASB1* gene

We analyzed the methylation status (β-values) of CpGs belonging to the *ASB* gene family between 8 ICM patients and 8 CNT donors using the 850K methylation array. Analysis of CpG differential methylation revealed the presence of only one hypermethylated CpG site of all *ASB* family, located in the *ASB1* gene (chr2:239344401-239344627) with a $\Delta\beta>0.1$. The hypermethylated CpG site (cg11189868), displayed a differential methylation profile of $0.26\Delta\beta$, *P*<0.05 (Figure 1A). We also validated these results through pyrosequencing, observing a $0.23\Delta\beta$, *P*<0.05 (Figure 1B).

Further, we performed an analysis of *ASB1* mRNA levels through RNA-sequencing and we found a downregulation of *ASB1* gene expression of -1.2 fold, *P*<0.05 (Figure 1C).

**Relationships between *ASB1* differential methylation and LV function and performance**

We sought to investigate the potential relationships between *ASB1* differential methylations and expression and hemodynamic and echocardiographic parameters of ICM patients. Interestingly, the differential methylation pattern of *ASB1* cg11189868 was strongly linked to SV (*r* = -0.929, *P* = 0.001) and LVEF (*r* = -0.849, *P* = 0.008) (Figure 2). This *ASB1* methylation profile also related to end-systolic and end-diastolic LV diameters (*r* = -0.743, *P* = 0.035 for both).

**DISCUSSION**

In this study, we analyse the methylation profile of *ASB* gene family in ICM patients, showing the presence of a differentially methylated CpG site located at the *ASB1* gene. None of the other *ASB* family genes showed methylation changes. This analysis demonstrates the presence, not previously reported, of a strong association between a differentially methylated pattern, validated by pyrosequencing, of *ASB1* gene in ICM subjects with the hemodynamic status, LV performance and cardiac function of these patients.

In previous studies, we have analysed the transcriptomic changes in cytoskeletal components of HF patients,[15] showing important alterations and links with LV dysfunction. Importantly, the gene coding of ankyrin repeat domain 1, *ANKRD1*, showed relationships with functional status of these patients, indicating a relevant role of this ankyrin gene in HF.

*ASB* gene family codify Ankyrin repeat and SOCS box proteins, being involved in protein-protein interactions acting as adaptors that target proteins for proteasomal degradation.[18] Scant data are available about the specific function of *ASB1*, relating it with alterations in spermatogenesis,[23] moreover, no studies have been conducted in cardiac tissues, but its protein superfamily has relevant implications in controlling the skeletal muscle contractile apparatus structural fixation and adequate regulation of differentiation steps.[17] *ASB2* has been implicated as a negative regulator of skeletal muscle mass through the TGF-β pathway, indicating that increased levels prevents hypertrophy.[19] Our results and the structural domain similarity in this gene family suggest a similar function for *ASB1* in the heart muscle. As demonstrated, gain of methylation of *ASB1* CpG island closely relates to LV function, dimensions, and output, and none of the other 18 *ASB* family genes show such change, indicating that an increased degree of methylation may be an indicator of deteriorating hemodynamic and cardiac function. In contrast, the *ASB1* gene expression calculated by means of the RNA-sequencing technique does not showed any LV significant relationships, suggesting a prominent role for this DNA methylation, maybe related to an unknown specific function in coding.

## Conclusions

Our findings strongly link a specific *ASB1* methylation pattern to LV morphology and performance in end stage ICM, and provide new insight and raising questions regarding which is the functionally relevant genome for the ischaemic failing myocardium.

**SOURCE OF FUNDING**

This work was supported by the National Institute of Health "Fondo de Investigaciones Sanitarias del Instituto de Salud Carlos III" [PI13/00100; PI14/01506], CIBERCV [CB16/11/00261], the European Regional Development Fund (FEDER), and RETICS [12/0042/0003].

**ACKNOWLEDGMENTS**

The authors thank the Transplant Coordination Unit (University and Polytechnic Hospital La Fe) for their help in obtaining the samples.

**CONFLICTS OF INTEREST**

The authors declare no conflict of interests.

**REFERENCES**

1. Rosello-Lleti E, Tarazon E, Barderas MG, Ortega A, Molina-Navarro MM, Martinez A, Lago F, Martinez-Dolz L, Gonzalez-Juanatey JR, Salvador A, Portoles M, Rivera M. ATP synthase subunit alpha and LV mass in ischaemic human hearts. *J Cell Mol Med* 2015;**19**(2):442-51.

2. Ortega A, Tarazon E, Rosello-Lleti E, Gil-Cayuela C, Lago F, Gonzalez-Juanatey JR, Cinca J, Jorge E, Martinez-Dolz L, Portoles M, Rivera M. Patients with Dilated Cardiomyopathy and Sustained Monomorphic Ventricular Tachycardia Show Up-Regulation of KCNN3 and KCNJ2 Genes and CACNG8-Linked Left Ventricular Dysfunction. *PLoS One* 2015;**10**(12):e0145518.

3. Hashimoto Y, Zumwalt TJ, Goel A. DNA methylation patterns as noninvasive biomarkers and targets of epigenetic therapies in colorectal cancer. *Epigenomics* 2016.

4. Fasanelli F, Baglietto L, Ponzi E, Guida F, Campanella G, Johansson M, Grankvist K, Johansson M, Assumma MB, Naccarati A, Chadeau-Hyam M, Ala U, Faltus C, Kaaks R, Risch A, De Stavola B, Hodge A, Giles GG, Southey MC, Relton CL, Haycock PC, Lund E, Polidoro S, Sandanger TM, Severi G, Vineis P. Hypomethylation of smoking-related genes is associated with future lung cancer in four prospective cohorts. *Nat Commun* 2015;**6**:10192.

5. Wilhelm T, Lipka DB, Witte T, Wierzbinska JA, Fluhr S, Helf M, Mucke O, Claus R, Konermann C, Nollke P, Niemeyer CM, Flotho C, Plass C. Epigenetic silencing of AKAP12 in juvenile myelomonocytic leukemia. *Epigenetics* 2016;**11**(2):110-9.

6. Yu L, Chibnik LB, Srivastava GP, Pochet N, Yang J, Xu J, Kozubek J, Obholzer N, Leurgans SE, Schneider JA, Meissner A, De Jager PL, Bennett DA. Association of Brain DNA methylation in SORL1, ABCA7, HLA-DRB5, SLC24A4, and BIN1 with pathological diagnosis of Alzheimer disease. *JAMA Neurol* 2015;**72**(1):15-24.

7. Singh SM, O'Reilly R. (Epi)genomics and neurodevelopment in schizophrenia: monozygotic twins discordant for schizophrenia augment the search for disease-related (epi)genomic alterations. *Genome* 2009;**52**(1):8-19.

8. Kathirvel M, Mahadevan S. The role of epigenetics in tuberculosis infection. *Epigenomics* 2016;**8**(4):537-49.

9. Castro R, Rivera I, Struys EA, Jansen EE, Ravasco P, Camilo ME, Blom HJ, Jakobs C, Tavares de Almeida I. Increased homocysteine and S-adenosylhomocysteine concentrations and DNA hypomethylation in vascular disease. *Clin Chem* 2003;**49**(8):1292-6.

10. Chen J, Yang T, Yu H, Sun K, Shi Y, Song W, Bai Y, Wang X, Lou K, Song Y, Zhang Y, Hui R. A functional variant in the 3'-UTR of angiopoietin-1 might reduce stroke risk by interfering with the binding efficiency of microRNA 211. *Hum Mol Genet* 2010;**19**(12):2524-33.

11. Volkmann I, Kumarswamy R, Pfaff N, Fiedler J, Dangwal S, Holzmann A, Batkai S, Geffers R, Lother A, Hein L, Thum T. MicroRNA-mediated epigenetic silencing of sirtuin1 contributes to impaired angiogenic responses. *Circ Res* 2013;**113**(8):997-1003.

12. Moran S, Arribas C, Esteller M. Validation of a DNA methylation microarray for 850,000 CpG sites of the human genome enriched in enhancer sequences. *Epigenomics* 2016;**8**(3):389-99.

13. Rodriguez-Penas D, Feijoo-Bandin S, Garcia-Rua V, Mosquera-Leal A, Duran D, Varela A, Portoles M, Rosello-Lleti E, Rivera M, Dieguez C, Gualillo O, Gonzalez-Juanatey JR, Lago F. The Adipokine Chemerin Induces Apoptosis in Cardiomyocytes. *Cell Physiol Biochem* 2015;**37**(1):176-92.

14. Ortega A, Gil-Cayuela C, Tarazon E, Garcia-Manzanares M, Montero JA, Cinca J, Portoles M, Rivera M, Rosello-Lleti E. New Cell Adhesion Molecules in Human Ischemic Cardiomyopathy. PCDHGA3 Implications in Decreased Stroke Volume and Ventricular Dysfunction. *PLoS One* 2016;**11**(7):e0160168.

15. Herrer I, Rosello-Lleti E, Rivera M, Molina-Navarro MM, Tarazon E, Ortega A, Martinez-Dolz L, Trivino JC, Lago F, Gonzalez-Juanatey JR, Bertomeu V, Montero JA, Portoles M. RNA-sequencing analysis reveals new alterations in cardiomyocyte cytoskeletal genes in patients with heart failure. *Lab Invest* 2014;**94**(6):645-53.

16. McDaneld TG, Spurlock DM. Ankyrin repeat and suppressor of cytokine signaling (SOCS) box-containing protein (ASB) 15 alters differentiation of mouse C2C12 myoblasts and phosphorylation of mitogen-activated protein kinase and Akt. *J Anim Sci* 2008;**86**(11):2897-902.

17. Tee JM, Peppelenbosch MP. Anchoring skeletal muscle development and disease: the role of ankyrin repeat domain containing proteins in muscle physiology. *Crit Rev Biochem Mol Biol* 2010;**45**(4):318-30.

18. Nie L, Zhao Y, Wu W, Yang YZ, Wang HC, Sun XH. Notch-induced Asb2 expression promotes protein ubiquitination by forming non-canonical E3 ligase complexes. *Cell Res* 2011;**21**(5):754-69.

19. Davey JR, Watt KI, Parker BL, Chaudhuri R, Ryall JG, Cunningham L, Qian H, Sartorelli V, Sandri M, Chamberlain J, James DE, Gregorevic P. Integrated expression analysis of muscle hypertrophy identifies Asb2 as a negative regulator of muscle mass. *JCI Insight* 2016;**1**(5).

20. Ponikowski P, Voors AA, Anker SD, Bueno H, Cleland JG, Coats AJ, Falk V, Gonzalez-Juanatey JR, Harjola VP, Jankowska EA, Jessup M, Linde C, Nihoyannopoulos P, Parissis JT, Pieske B, Riley JP, Rosano GM, Ruilope LM, Ruschitzka F, Rutten FH, van der Meer P, Authors/Task Force M. 2016 ESC Guidelines for the diagnosis and treatment of acute and chronic heart failure: The Task Force for the diagnosis and treatment of acute and chronic heart failure of the European Society of Cardiology (ESC)Developed with the special contribution of the Heart Failure Association (HFA) of the ESC. *Eur Heart J* 2016;**37**(27):2129-200.

21. Macrae DJ. The Council for International Organizations and Medical Sciences (CIOMS) guidelines on ethics of clinical trials. *Proc Am Thorac Soc* 2007;**4**(2):176-8, discussion 178-9.

22. Edgar R, Domrachev M, Lash AE. Gene Expression Omnibus: NCBI gene expression and hybridization array data repository. *Nucleic Acids Res* 2002;**30**(1):207-10.

23. Kile BT, Metcalf D, Mifsud S, DiRago L, Nicola NA, Hilton DJ, Alexander WS. Functional analysis of Asb-1 using genetic modification in mice. *Mol Cell Biol* 2001;**21**(18):6189-97.

**Table 1. Clinical characteristics of ischaemic cardiomyopathy patients.**

|  | ICM (n=8) | ICM (n=13) | ICM (n=14) |
|---|---|---|---|
|  | **Epigenomics** | **RNA sequencing** | **Pyrosequencing** |
| Age (years) | 53±5 | 54±7 | 53±6 |
| Gender male (%) | 100 | 100 | 100 |
| NYHA class | III-IV | III-IV | III-IV |
| BMI (kg/m$^2$) | 28±3 | 26±4 | 28±4 |
| Haemoglobin (mg/dL) | 14±2 | 14±3 | 14±2 |
| Haematocrit (%) | 44±4 | 41±6 | 42±5 |
| Total cholesterol (mg/dL) | 152±43 | 162±41 | 171±46 |
| Prior hypertension (%) | 25 | 30 | 39 |
| Prior smoking (%) | 88 | 84 | 92 |
| Diabetes mellitus (%) | 63 | 38 | 54 |
| LVEF (%) | 24±6 | 24±4 | 23±5 |
| LVESD (mm) | 57±8 | 55±7 | 56±7 |
| LVEDD (mm) | 65±7 | 64±7 | 64±7 |

ICM, ischaemic cardiomyopathy; NYHA, New York Heart Association; BMI, body mass index; LVEF, ejection fraction; LVESD, left ventricular end-systolic diameter; LVEDD, left ventricular end-diastolic diameter.

**LEGENDS**

**Figure 1. Differentially methylated profile of *ASB1* and gene expression in ICM patients. A.** Methylation pattern of the *ASB1* gene in ICM patients showing the expansion of the differentially methylated CpG sites between ICM and CNT. **B**. Validation of DNA methylation CpG island by pyrosequencing. **C.** Gene expression analysis of *ASB1* gene through RNA-sequencing. CNT, control; ICM, ischaemic cardiomyopathy; TSS, transcription start site. *P<0.05.

**Figure 2. Relationships of the differentially methylated CpG site with the invasive-calculated SV and with eco-Doppler based EF.** CNT, control; EF, ejection fraction; ICM, ischaemic cardiomyopathy; SV, stroke volume.

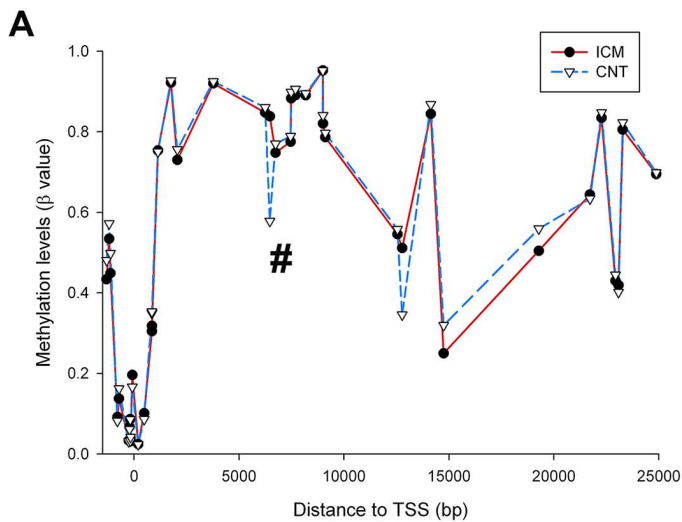
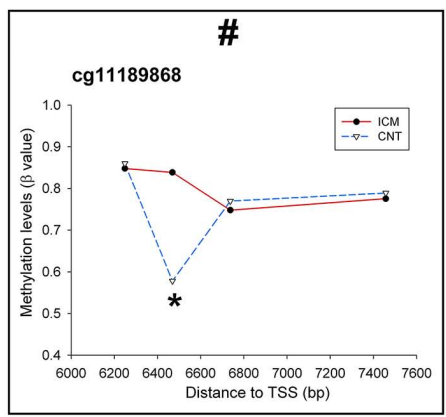
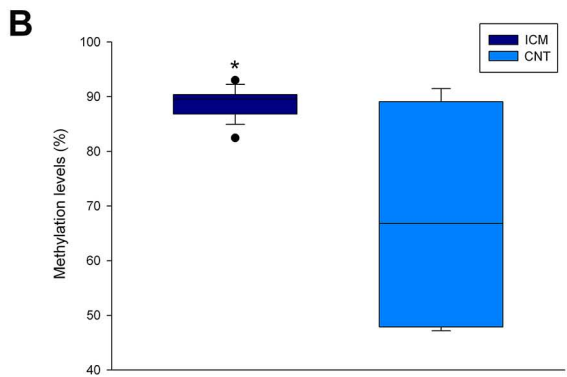
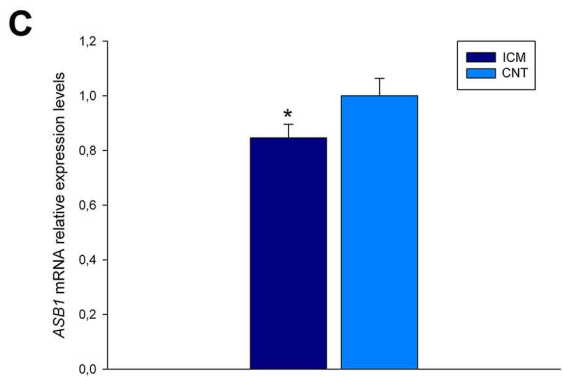

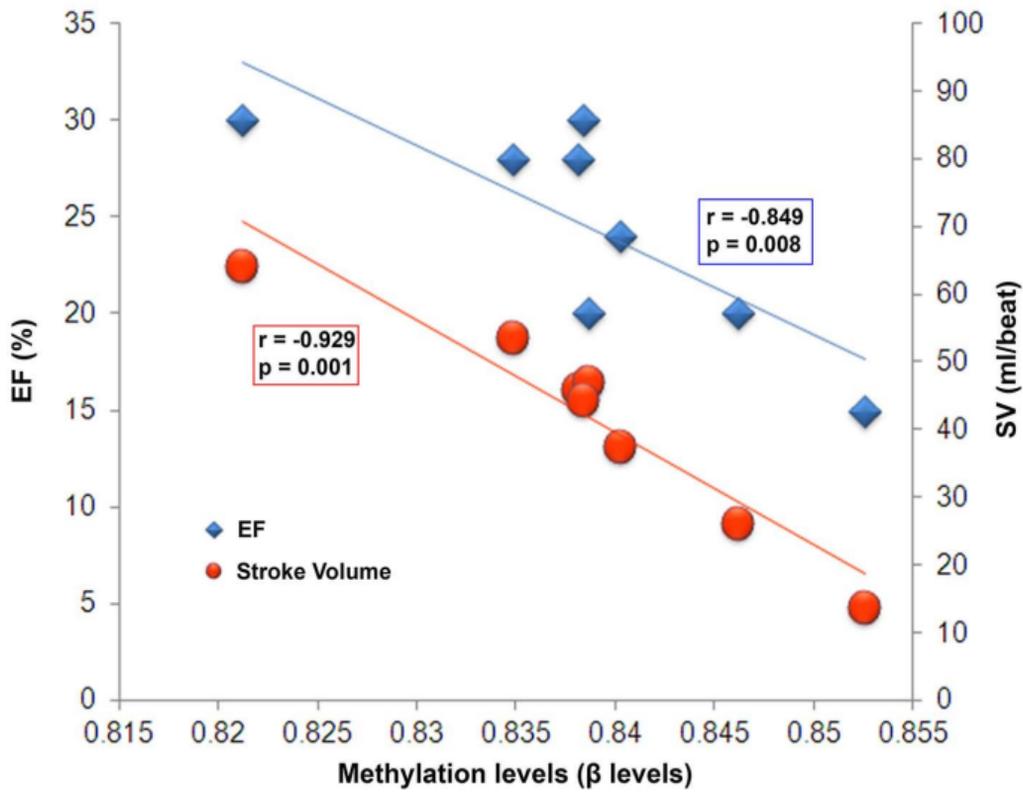